\documentclass[times]{tum-article}

\usepackage{cite}
\usepackage{multirow}

\newenvironment{adjustwidth}[2]{}{}

\title{bertha: Project Skeleton for Scientific Software}
\author{Michael Riesch\orcid{0000-0002-4030-2818}\authormark{\Letter},
  Tien Dat Nguyen\orcid{0000-0003-2522-7335}, and
  Christian Jirauschek\orcid{0000-0003-0785-5530}}
\affil{Department of Electrical and Computer Engineering, Technical
  University of Munich, Arcisstr. 21, 80333 Munich, Germany}
\email{michael.riesch@tum.de}
\date{Received: 10 December 2019 / Accepted: 04 March 2020 / Published:
  23 March 2020
  \thanks{This is a post-peer-review, pre-copyedit version of an open access
    article published in PLOS ONE under the terms of the Creative Commons
    Attribution License, which permits unrestricted use, distribution, and
    reproduction in any medium, provided the original author and source are
    credited. The final authenticated version is available online at:
    \url{https://dx.doi.org/10.1371/journal.pone.0230557}}}

\begin{document}

\maketitle

\begin{abstract}
Science depends heavily on reliable and easy-to-use software packages, such
as mathematical libraries or data analysis tools. Developing such packages
requires a lot of effort, which is too often avoided due to the lack of
funding or recognition. In order to reduce the efforts required to create
sustainable software packages, we present a project skeleton that ensures
the best software engineering practices from the start of a project, or
serves as reference for existing projects.

\end{abstract}

\section{Introduction}

In a recent essay in Nature~\cite{nowogrodzki2019}, a familiar dilemma in
science was addressed. On the one hand, science relies heavily on
open-source software packages, such as libraries for mathematical operations,
implementations of numerical methods, or data analysis tools. As a
consequence, those software packages need to work reliably and should be
easy to use. On the other hand, scientific software is notoriously
underfunded and the required efforts are achieved as side projects or by the
scientists working in their spare time.

Indeed, a lot of effort has to be invested beyond the work on the actual
implementation -- which is typically a formidable challenge on its own. This
becomes apparent from literature on software engineering in general (such as
the influential ``Pragmatic Programmer''~\cite{hunt1999pragmatic}), and in
scientific contexts in particular (e.g.,
\cite{prlic2012, bangerth2013, wilson2014, wilson2017}). The vast number of
best practices guides and development guidelines available (e.g., those
published by the German Aerospace Center (DLR)~\cite{schlauch2018dlrguide}
and the Netherlands eScience Center~\cite{netherlands-guide2019}) further
underlines the importance of the topic and may serve as guidance, but often
scientists lack the time and/or formal training in software engineering
required to ensure sustainable software
development~\cite{nowogrodzki2019, wilson2014, wilson2017}. Too often, this
results in poorly maintained software projects of questionable reliability
and usability.

Given all this, once again the goal is to achieve much with little effort.
Therefore, in this paper we present a project skeleton that may serve as
solid yet lightweight base for a small to medium-scale scientific software
project. In the use case we envisage, scientists can create an instance of
this template in just a few clicks. This instance implements essential best
practices in software engineering from the very start. After performing a
minimal number of customizations, the scientist can soon start working on the
actual implementation and can concentrate on what really matters. In the
context of pure C++~\cite{kracejic2015} or Python~\cite{ioannides2018}
projects, such skeletons have already proven their value. In the scope of
this work, we focus on scientific software libraries which are written in the
C++ programming language for performance reasons and feature bindings for
Python in order to provide an easy-to-use interface to the user. These
programming languages are widely accepted in both open-source and
high-performance computing (HPC) communities, and should therefore be
considered a reasonable choice. To the best of our knowledge, a project
skeleton for this combination of programming languages has not yet been
published.

It should be noted that the skeleton does not (and should not) cover every
eventuality (e.g., when support for the Fortran programming language is
required) but concentrates on one particular use case. This is contrary to
the recommendations in related literature, which are kept general and
language-agnostic on purpose. The rationale behind this decision is to keep
the template lightweight and avoid cluttering.

The paper at hand is organized as follows: In Section~\ref{sec:measures}, we
identify the essential best practices that are required to ensure
high-quality scientific software based on related literature and our own
experiences with our software projects (e.g., the mbsolve
software~\cite{riesch2018oqel,mbsolve-github}, a solver for the generalized
Maxwell-Bloch equations~\cite{jirauschek2019ats}). Subsequently, we present
our project skeleton and discuss the specific implementation of the
measures identified in Section~\ref{sec:implementation}. As already stated
above, some minor customization steps are required.
Section~\ref{sec:customization} gives an overview of these steps and thereby
an introduction to the (potential) user. Finally, we conclude with a short
summary and give an outlook on future work, i.e., additional tools and
measures that further improve the quality of scientific software projects.

\section{Best practices in scientific software engineering}
\label{sec:measures}

This section describes the essential recommendations and best practices from
related literature~\cite{nowogrodzki2019, hunt1999pragmatic, prlic2012, %
  bangerth2013, wilson2014, wilson2017, schlauch2018dlrguide, %
  netherlands-guide2019} that serve as basis for the project skeleton.
All recommendations are language-agnostic and grouped into seven categories
with no particular order of importance. Table~\ref{tab:practices} gives an
overview of the best practices.

\begin{table}[tb]
  \centering
  \begin{adjustwidth}{-2.25in}{0in}
    \caption{Overview of best practices in software engineering for
      scientific software projects. For each best practice, implementation
      candidates are listed where the selected choice is denoted in bold.}
    \label{tab:practices}
    \begin{tabular}{lll}
      \hline
      \textbf{Group} & \textbf{Best practice} &
      \textbf{Implementation candidates (not exhaustive)}\\
      \hline
      \multirow{3}{*}{Project management}  & Version control system &
      \textbf{git}, mercurial, svn\\
      \cline{2-3}
      & Project management tool      &
      \textbf{GitLab}, \textbf{GitHub}, Bitbucket, JIRA\\
      \cline{2-3}
      & Workflow & \textbf{GitLab Flow}, GitHub Flow, git flow\\
      \hline
      \multirow{3}{*}{Coding style} & Code formatting style &
      \textbf{Mozilla}, LLVM, Google, Chromium\\
      \cline{2-3}
      & Code formatting tool & \textbf{clang-format}\\
      \cline{2-3}
      & Static code analysis & clang-tidy, cppcheck, cpplint\\
      \hline
      \multirow{2}{*}{Independence} & Use open file formats &
      e.g., JSON, CSV, HDF5\\
      \cline{2-3}
      & Use open-source libraries &
      e.g., Eigen, FFTW, GNU Scientific Library\\
      \hline
      \multirow{2}{*}{Automation} & Continuous integration &
      \textbf{gitlab-ci}, \textbf{Travis CI}, AppVeyor, Microsoft Azure\\
      \cline{2-3}
      & Build automation &
      \textbf{CMake}, GNU make, Bazel, Ninja, MS Build\\
      \hline
      \multirow{2}{*}{Documentation} & Function reference &
      \textbf{Doxygen}, Sphinx (with Breathe)\\
      \cline{2-3}
      & ``Big picture'' documentation &
      \textbf{Markdown}, reStructuredText\\
      \hline
      \multirow{2}{*}{Testing} & Unit test framework &
      \textbf{Catch2}, Google Test, Boost Test Library\\
      \cline{2-3}
      & Code coverage report &
      \textbf{gcov}, various commercial tools\\
      \hline
      \multirow{2}{*}{Deployment} & Package binaries &
      \textbf{conda}, Conan, Debian apt\\
      \cline{2-3}
      & Online documentation &
      \textbf{GitLab Pages}, \textbf{GitHub Pages}, readthedocs.io\\
      \hline
    \end{tabular}
  \end{adjustwidth}
\end{table}

\subsection{Project management}

Most software projects in a scientific context start with a single developer.
However, over time the projects are likely to grow, be extended, and possibly
taken over by other developers. Building a developer community is crucial for
the success of the project in general and in particular for open-source
projects~\cite{bangerth2013}. Therefore, the project infrastructure should be
able to handle multiple developers from the very start.

All of the guidelines in literature we have found mention the usage of a
version control system (VCS). This is beneficial even for the single
developer, as a VCS intrinsically features a backup solution and
synchronization between different machines. Once more developers start
working on the project, this enables transparent collaboration. By using a
VCS, the ``Make Incremental Changes'' paradigm~\cite{wilson2014, wilson2017}
can be implemented easily and the intrinsically generated development history
may serve as rudimentary documentation of design
decisions~\cite{bangerth2013}.

In a more advanced scenario, the VCS is coupled with a project management
tool that provides a means of communication within the developer team, and
thereby further enhances transparency. As the communication logs are
available for developers who join the team at a later stage, this also
provides a certain form of documentation~\cite{bangerth2013}. One essential
element of a project management tool is a ticket system or issue tracker.
Issues are requests for a certain change (such as a bug fix or feature
implementation) and play a crucial role in modern iterative and incremental
software development processes, such as feature-based
development~\cite{highsmith2010}. As the name suggests, issue trackers keep
track of issues from their creation (by users or developers) to their
completion in the form of an accepted solution by the
developer~\cite{schlauch2018dlrguide}. Modern project management tools also
include convenient mechanisms for code review. Similar to a scientific paper,
a rigorous review process may be time-intensive and annoying, but eventually
yields solutions of higher quality and wider acceptance~\cite{wilson2014}.

\subsection{Code quality}

Just as we care about language style when writing a scientific article, so
we should care about coding style when writing scientific software. Here,
we should bear the mottos
``Write Programs for People, Not Computers''~\cite{wilson2014} and
``Don't Repeat Yourself''~\cite{hunt1999pragmatic, wilson2014} in mind and
produce easily readable and modular code. In developer teams, it is crucial
to agree on a certain coding style at the beginning of the project. The
coding style usually consists of two parts: rules for formatting code and
best practices for programming in the respective language. Code formatting
tools enable manual and automated checks to establish whether the source code
is compliant with agreed code formatting rules~\cite{netherlands-guide2019}.
Analogously, static code analysis tools check whether the agreed best
practices are violated~\cite{schlauch2018dlrguide}.

\subsection{Independence}
\label{sec:independence}

Some guidelines recommend that open standards, protocols, and file formats
should be used wherever possible (e.g., the HDF5 format for large data
sets~\cite{netherlands-guide2019}). Thereby, vendor lock-in situations are
avoided which would arise, for example, if a certain source code can only be
compiled using a certain compiler brand or version. Our general
recommendation here is to provide solutions that work with the most widely
used operating systems and compilers (and possibly combinations thereof)
from the very start.

Following the advice that one should never reinvent the wheel, established
software libraries and tools are often used to speed up development
processes. Here, we recommend using open-source components unless there is a
strong reason not to. This is in agreement with the interoperability and
reusability part of the FAIR principle~\cite{wilkinson2016, lamprecht2019}.

\subsection{Automation}
\label{sec:automation}

We should
``Let the Computer Do the Work''~\cite{hunt1999pragmatic, wilson2014} and
automate repetitive tasks such as building the software, running tests,
performing quality checks, and deploying the generated artifacts (typically,
software in binary form and documentation) to a software repository.
Otherwise, those tedious tasks are most likely postponed, not done at all,
or performed only partially. Here, continuous integration (CI) tools are
helpful as different jobs can be defined and grouped into stages, which are
executed every time the developers push changes to the version control
repository. Then, the developers receive feedback on the changes, which is an
essential part of the ``Make Incremental Changes'' strategy~\cite{wilson2014}.

The feedback typically consists of (at least) two parts, which are briefly
outlined. First, the build process should run in an automated and
platform-independent fashion. Here, it is particularly important that
third-party dependencies are found without hard coded paths. The output of
the build process tells the developers whether the build on different
platforms was successful. This is especially beneficial as most developers
develop on a certain platform and the code is not intrinsically tested on
other platforms (different operating systems, different compiler versions,
etc.). Second, test programs can be executed automatically on different
platforms. For example, unit tests can help to verify the correct behavior
of certain functions or modules of the software. Functional tests, on the
other hand, help to gain more confidence in the overall function of the
software~\cite{schlauch2018dlrguide}.

It makes sense to define the continuous integration pipelines as early as
possible, so that the developers benefit from the feedback from the very
beginning. Thereby, bugs in the software (in particular regressions) can be
detected early. Furthermore, the effectiveness of optimizations can be
assessed while the correct operation of the software is ensured.

\subsection{Documentation}
\label{sec:documentation}

In order to make scientific software reusable, providing documentation to
users and developers is one of the most important
steps~\cite{nowogrodzki2019, hunt1999pragmatic, prlic2012, bangerth2013, %
  wilson2014, wilson2017, schlauch2018dlrguide, netherlands-guide2019}.
Bangerth and Heister~\cite{bangerth2013} list five items that the
documentation should contain: traditional comments, function level
documentation, class level documentation, overview of how modules interact,
and complete examples in tutorial form. As to traditional comments, it is
good practice to
``Document Design and Purpose, Not Mechanics''~\cite{wilson2014} and avoid
obvious comments. Function and class level documentation is typically
generated based on comments in code using special annotation. The resulting
reference manual is particularly interesting for developers and advanced
users who need to know the details. On the other hand, the module overview
documentation should inform new users about the big picture. This information
is typically written into the files README (aim of the software, installation
notes, list of dependencies), CHANGELOG (overview of releases, features,
known bugs), CODE\_OF\_CONDUCT and CONTRIBUTING (guidelines for (potential)
developers), as well as TUTORIAL (guide for (potential)
users)~\cite{wilson2017}.

\subsection{Testing}

As mistakes are natural and are bound to happen, we should plan for them and
develop strategies on how to detect them as early as
possible~\cite{wilson2014}. Automated testing, the importance of which has
already been underlined in Section~\ref{sec:automation}, is the cornerstone
of such strategies. It should be noted that the effectiveness of tests should
be monitored as well. Here, code coverage tools are useful as they are able
to detect code parts which are not covered by the executed
tests~\cite{schlauch2018dlrguide}.

Again, we stress that certain measures, such as writing unit tests, should be
carried out from the very beginning. Apart from their use in automated
testing, unit tests may have a positive effect on the code design. Since
modular code is usually testable, performing unit tests can be considered a
necessary requirement for modular code~\cite{hunt1999pragmatic}.

\subsection{Deployment}

Whether or not a certain software project is used depends to a large degree
on the ability to distribute it~\cite{bangerth2013}. Hence, it is
advisable to package the software and distribute it using an established
software repository~\cite{nowogrodzki2019}. Similar to the practices
discussed above, it is important that the deployment is carried out
automatically and as early as possible~\cite{prlic2012}.

\section{Implementation of the project skeleton}
\label{sec:implementation}

\begin{figure}[tb]
  \begin{adjustwidth}{-2.25in}{0in}
    \centering
    \includegraphics{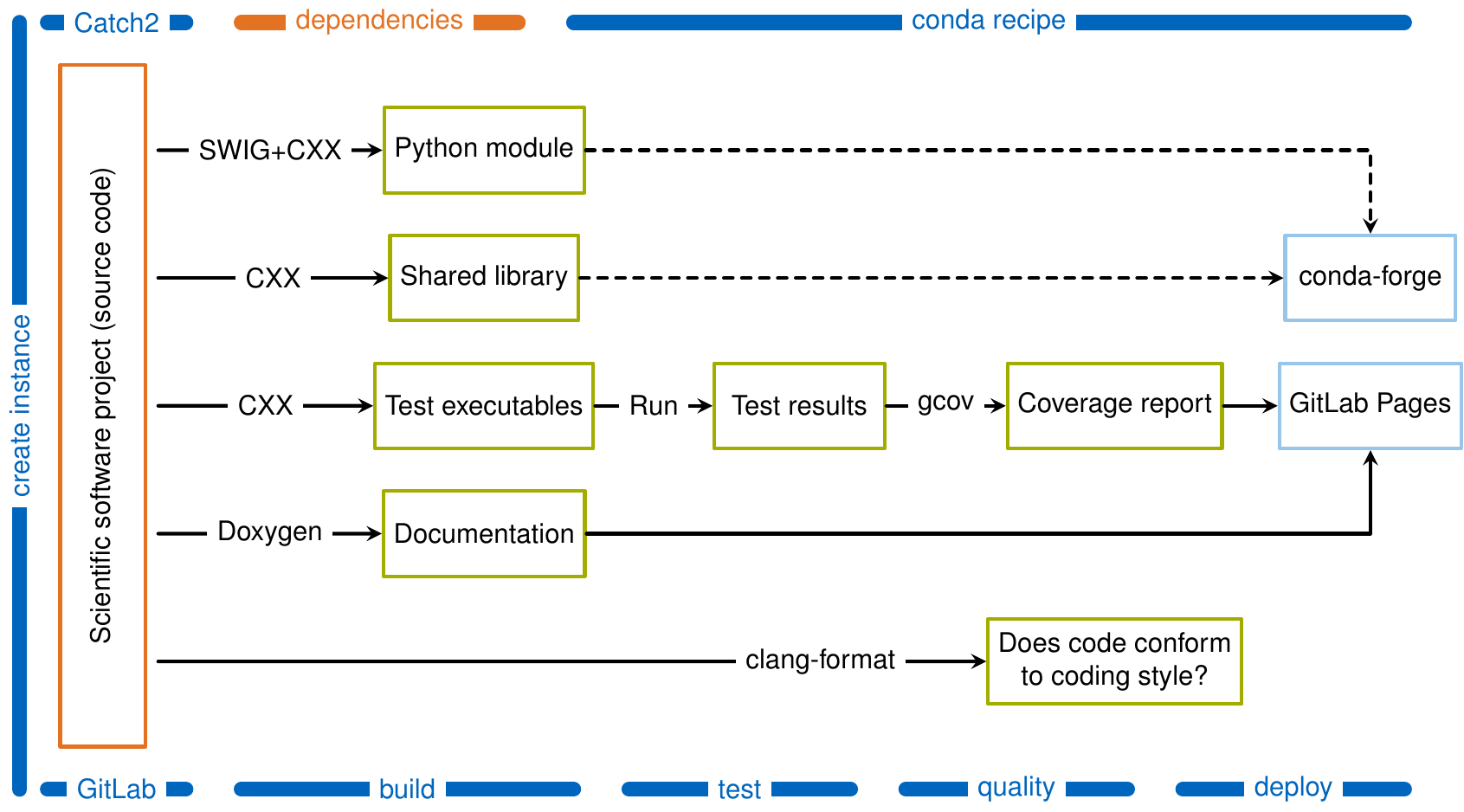}
    \caption{Overview of the project skeleton. The source code and
      dependencies of a scientific software project are denoted in orange.
      These are the parts the developer has to provide. The presented
      skeleton guides the project from creation to deployment. Here, the
      arrows denote jobs that are created by the CMake build system. These
      jobs are triggered during the different continuous integration stages
      (build, tests, quality, deploy) or (in the case of the dashed arrows)
      by the conda-forge build service that follows the
      recipe~\cite{bertha-feedstock}.
      The job names indicate the tools in use, where CXX represents one of
      the C++ compilers that are supported by CMake.}
    \label{fig:overview}
  \end{adjustwidth}
\end{figure}

Based on the (general and language-agnostic) best practices introduced in
the section above, we implement measures for a C++ software library with
bindings for the Python language in this section. The result is publicly
available~\cite{bertha} and may serve as a template for new projects or
reference for existing projects. Figure~\ref{fig:overview} outlines the
skeleton approach.

It should be noted that there may be different ways to implement a certain
measure. For the sake of simplicity, we discuss only one or two
possibilities for most measures. Following the recommendations in
Section~\ref{sec:independence}, we have selected open-source tools and
libraries exclusively. Thereby, one particular lightweight solution is
provided for scientists who are new to the topic, while the advanced users
may replace a certain implementation of a measure with another library or
tool of their choice.

Since a project skeleton does not include a real implementation, best
practices regarding planning, structuring, and writing code can hardly be
demonstrated. In this regard, we refer the reader to available literature on
the topic, such as~\cite{hunt1999pragmatic}, and focus on the project
skeleton that provides the required infrastructure.

\subsection{Usage of a version control system (VCS) and appropriate workflow}

A multitude of version control systems has been published and used over the
last three decades. We stick to our criterion that the software must be
open-source and note that git has received much attention since it was first
released in 2005. It features distributed version control and a flexible
branching model, rendering it perfectly suited for open-source projects.
However, the flexible branching model might, at the same time, be a
significant drawback. Each project should define a workflow to show how
changes are developed, tested, and integrated. As usual, it makes sense to
use something established, such as the GitLab Flow~\cite{gitlab-flow}. This
workflow uses feature branches to develop and test new features or bug fixes.
Once the changes on the feature branch fulfill the requirements and pass the
automated tests and quality checks, the developer can open a merge request.
A maintainer can subsequently merge the changes in the main development
branch. Additionally, the GitLab Flow allows stable branches and different
environments (such as production) in which further restrictions may apply.
The latter features are not required at the initial stage of a project, but
underline that the GitLab Flow is simple enough for small projects yet
powerful enough for large and established projects.

\subsection{Usage of a project management tool including issue tracking}

There are several management tools and hosting platforms that can be combined
with the git version control system with different strengths and drawbacks.
Here, we would like to leave the choice to the developers and provide two
possible solutions for the undecided.

Over the last decade, the GitHub platform has received significant attention.
It provides free public git repositories and integrations with other services
(such as the zenodo repository for storing research output). Due to its
prominence, we have decided to provide a mirror repository of the project
skeleton in GitHub~\cite{bertha-github}.
This repository is marked as project template, which allows a new project to
be instantiated with a few clicks. As to continuous integration, GitHub
offers support for external CI providers such as Travis CI, AppVeyor, and
Microsoft Azure. These services are typically free for open-source projects
and configured using a YAML file, where CI jobs can be described. As an
example, we have added a basic configuration file for Travis CI that triggers
build and unit tests on Linux, Windows, and macOS platforms given that the
user registers on Travis CI, where the CI operation can be activated for the
repository in question.

Alternatively, the GitLab platform can be used, which is conceptually similar
to GitHub, the main difference being the possibility of self-hosting the
platform on a local server for free. While the concepts (such as Pages and
Releases) are similar, there are slight differences. For example, the project
template instantiation mechanism is different. At this point, it is not
possible to create an instance of the project skeleton with a single click.
However, we aim to provide that feature in the near future~\cite{bertha-mr}.

GitLab.com provides free hosting and internal continuous integration services
for open-source projects. Currently, those internal CI services are restricted
to the Linux operating system. It is possible, however, to install GitLab's
CI suite on a local machine and connect it to GitLab. Alternatively, an
external service can be used for Windows or macOS operating systems. In the
event that the project should not be open-source, the self-hosted operation
mode may be selected. Here, the CI suite must be installed on local machines,
which can subsequently be connected to the local GitLab installation.

It should be noted that we did not add configuration files for all options to
the template in order to provide a lightweight skeleton. Instead, we included
the configuration file for the GitLab internal CI, which calls the targets
generated by the build systems. From this configuration file, corresponding
files for other CI services can be derived.

\subsection{Automated build system}

In particular when the C++ programming language is involved, the CMake
project provides well-established tools to build, test, and package software.
The main advantage of CMake (compared to alternatives such as GNU make,
Visual Studio, or Eclipse) is that a level of abstraction is introduced. The
configuration files consist of directives such as \texttt{add\_library} or
\texttt{find\_package} and are, therefore, quite easy to read and understand.
Based on those configuration files, project files for the aforementioned
alternatives (and many other build systems) can be generated. Thereby, the
software project can be built for different operating systems or using
different compilers. In addition, CMake features a mechanism for finding
third-party libraries and tools. This feature is essential for
cross-platform dependency management.

As a proof of concept, we have added a simple shared library written in C++
to our project skeleton. It features a simple class \texttt{device} with two
member variables that represent its start and end coordinates, respectively.
An instance of this class can be created using one of two constructors, where
either the coordinates are specified directly, or the length can be set and
the start coordinate is assumed to be at the origin. Finally, a method
returns the length of the device.

For such a shared library, Python bindings can be generated conveniently
using the SWIG project. It is fully supported by CMake and requires only a
minimal configuration file, which basically specifies which C++ header files
should be considered when the interface is created. SWIG scans the specified
header files and automatically generates a Python module, which can be
subsequently imported and used in a Python project.

\subsection{Unit testing}

Ideally, the software is designed so that each unit of software (e.g., a
function) fulfills a certain, unique task (``Design by Contract''
technique~\cite{hunt1999pragmatic}). Furthermore, the implementation of each
unit is flawless. While the first goal can be achieved by careful design and
refactoring, the second statement is rarely true. As mentioned above,
mistakes will happen and we have to test whether the implementations of each
unit work correctly.

In the case of our simple C++ library, we have to check, for instance,
whether the calculation of the length yields the correct result. This can be
achieved by writing a unit test that creates an instance of the
\texttt{device} class, calls its \texttt{get\_length} method, and compares
the result of the method to the expected value. Also, whenever the user
specifies input data, the implementation should check whether those values
are reasonable and deal with invalid values (most likely, by throwing an
exception). Error handling code must be tested as well, for example by
creating a unit test in which the error is provoked on purpose and checking
whether the error handling code yields the correct behavior. As the number
of unit tests is expected to be large for a real life project, it is
recommended to use a unit test framework.

We chose the Catch2 library as it is open-source, lightweight and header-only.
Based on this library, we added a test executable with several unit tests to
our CMake build system. Here, we could rely on the CTest functionality of the
CMake project. Whether or not the unit tests cover all possible situations
can be assessed using code coverage tools. We have added the possibility of
using the gcov tool to the project skeleton. This tool generates profiling
information during the execution of tests. This information can be
subsequently converted to a human-readable report, in which metrics such as
line coverage are given on a per-file basis.

\subsection{Automatic code formatting}

Here, the clang-format tool constitutes a helpful and versatile instrument.
It can be configured using a single file, in which the code formatting rules
are specified. There are several predefined styles that can be used as-is,
or alternatively serve as a basis. It is also possible to define a certain
style from scratch, but we recommend using an existent style (with slight
modifications, if required).

In our project skeleton, the clang-format tool is integrated into the CMake
build system, making it easy for the user to format all source files
automatically. This functionality is also used to check whether the source
code conforms to the specified style in the scope of continuous integration.

\subsection{Documentation generation}

From the implementation point of view, we can separate the different types of
documentation listed in Section~\ref{sec:documentation} into two groups,
namely the function reference and the overview documentation. The function
reference is based on comments in the source code that use special annotation.
The information in those comments can be extracted using the Doxygen tool.
For the overview documentation, which provides the ``big picture'', it makes
sense to use a structured text format. Since Doxygen supports the Markdown
language, we chose to write files such as README.md and CONTRIBUTING.md in
this annotation. Both the overview documentation and function reference are
then transformed into static HTML pages that can be viewed locally or
uploaded to a web server.

We note that while Doxygen provides unchallenged support for in-source C++
documentation, the design of the generated HTML files appears a bit dated.
More advanced workflows are available that use Doxygen as input parser and
alternative tools to generate the static HTML pages. However, this is beyond
the scope of the work at hand.

\subsection{Automated packaging and deployment to a public repository}

While many operating systems or programming languages feature a common
repository for exchanging programs and libraries in binary form, it would be
beneficial to have a language-agnostic repository that covers all operating
systems. Fortunately, the conda system provides exactly this. Once a software
project is in a stable state, a recipe can be created on conda-forge that
defines the source of the project, the steps required to build it, and meta
information such as the name of the responsible maintainer. Based on this
recipe, the conda-forge build system automatically generates the binaries for
different platforms. Then, on each platform the resulting package can be
easily installed within a conda environment.

Most likely, the package has dependencies on other libraries. The conda
system offers a vast number of third-party components and convenient methods
of installing them. The environment approach already mentioned has a positive
effect on the dependency management, as in Windows it is generally impossible
to distinguish between different versions of a library (at least when
considering unmanaged C++ code), dubbed the ``DLL Hell''. Using conda
environments, however, it is possible to separate different versions in a
clean and convenient way.

The documentation generated could be included in a conda package as well.
However, we found it more appropriate to publish it on a web server for
visibility reasons. Both GitHub and GitLab offer the possibility of hosting
static HTML pages, such as those generated by Doxygen. With a few lines of
CI configuration, the documentation is automatically generated and uploaded.
See~\cite{bertha-doc} for an example.

\section{Creating a skeleton instance}
\label{sec:customization}

\begin{figure}
  \centering
  \begin{adjustwidth}{-2.25in}{0in}
    \includegraphics[width=\linewidth]{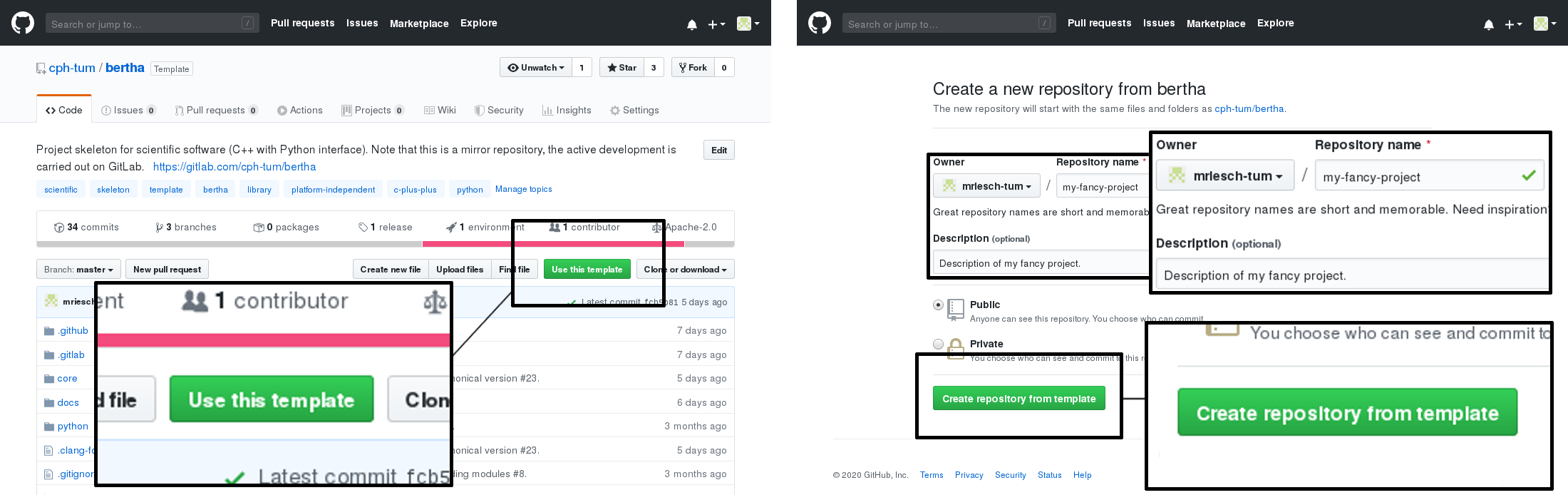}
    \caption{Creating an instance of the project skeleton on GitHub. On the
      project page of bertha~\cite{bertha-github}, click on
      ``Use this template''. In the following, enter the desired owner,
      repository name, and project description. The button
      ``Create repository from template'' will then create the instance.}
    \label{fig:screen-github}
  \end{adjustwidth}
\end{figure}

\begin{figure}
  \centering
  \begin{adjustwidth}{-2.25in}{0in}
    \includegraphics[width=\linewidth]{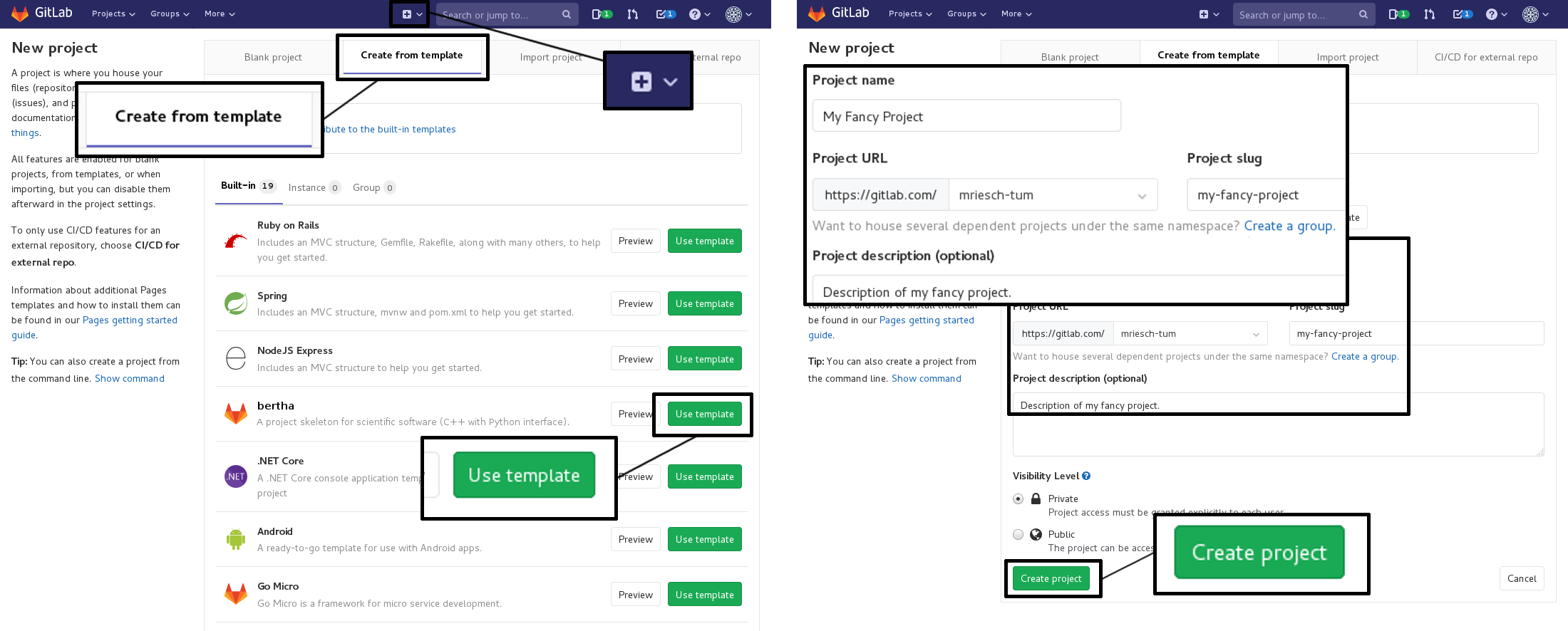}
    \caption{Creating an instance of the project skeleton in GitLab.
      Click on the plus button to create a new project. After selecting the
      ``Create from template'' tab, choose bertha by clicking
      ``Use this template'' (currently in development, see~\cite{bertha-mr}).
      Then, enter the project name and description and click
      ``Create project''.}
    \label{fig:screen-gitlab}
  \end{adjustwidth}
\end{figure}

In order to create a new project, the project skeleton can be cloned using
the mechanisms of either GitHub (see Fig.~\ref{fig:screen-github}) or
GitLab (as described in Fig.~\ref{fig:screen-gitlab}). Alternatively, the
files can be copied manually and added to a new repository. After the cloning
procedure, the skeleton can be adjusted to the needs of the new project.
The steps recommended and required are outlined briefly in the following.
For more detailed instructions, please refer to the ``Tutorial'' section in
the bertha documentation~\cite{bertha-doc}. Please note that registration on
Travis CI and activation of the project is required for continuous
integration support in GitHub.

\subsection{Setup stage}

At the beginning, it is important to define a meaningful name for the project
and replace bertha with this name throughout the project (e.g., in the CMake
build structure). Ensure that the name is not already used (e.g., in
conda-forge) if the project is to be open-source. Then, the project team
should agree on where to host the project (for internal use only or publicly
available), on the license for the project, and on the workflow. The latter
includes mainly the coding style and the version control workflow. Both
should be documented as soon as possible.

\subsection{Implementation stage}

At this point, the software project has a solid initial state. Now it is
time to add functionality. Here, consider writing the documentation first
(the contract), then implementing the functionality, and at the same time
writing unit tests. This approach will seem slow but improves the quality
of the design and helps to detect mistakes early on. Also, the CMake build
structure can be adjusted to add requirements (e.g., software libraries) or
additional modules (besides the existing core library).

\subsection{Publication stage}

In the case of an open-source project, the code should be distributed and
communicated as soon as it has some first functionality. For the
distribution of the project in binary form, the conda recipe for
bertha~\cite{bertha-feedstock} may serve as reference.

\section{Conclusion}

In the work at hand, we have presented a skeleton for scientific software
projects which consist of libraries written in the C++ programming language
and feature a Python interface. The skeleton contains the essential
elements required to ensure best software engineering practices. With this,
we hope to provide the scientific community with a helpful tool that saves
time during the setup of a new project. Based on our experience gained
during the development of the skeleton, creating a bertha instance may
replace at least one person month of evaluating tools, reading documentation,
and searching for answers in the internet.

Furthermore, this contribution may serve as checklist and reference for
existing projects. We hope that in both use cases -- building a project
from scratch and adapting an existing one -- the project skeleton will aid
the implementation of good practices in scientific software engineering and
consequently improve the quality and reusability of scientific software
projects.
At the moment, two of our simulation software projects are based on bertha.
The mbsolve software mentioned above is an open-source project in which
bertha is used to adapt the existing project. On the other hand, we have
created the first instance of bertha for a solver tool that is dedicated to
the numerical modeling of rapidly wavelength-swept Fourier domain mode-locked
fiber lasers~\cite{jirauschek2015fdml,jirauschek2017josab}. This tool is
developed in-house, which is allowed by the permissive license of the bertha
project. As we are going to use bertha in further simulation projects, we
plan to maintain it in the future. Nevertheless, we hope that in the near
future a community will form around the project skeleton that will use,
maintain, and extend bertha.

As a next step, the implementation of further measures is envisaged. For
example, a static code analysis tool could further improve the quality of
the code. Also, the generated documentation and quality reports should be
presented with a modern appearance. Finally, the project skeleton concept
can be transformed to other project classes in scientific software
engineering, such as a combination of a Fortran library with a Python
interface.

\section*{Acknowledgments}

The authors would like to thank Christian Widmann and Michael Haider from our
group, as well as Alexander Valavanis, Carina Haupt, Joachim Wuttke, Richard
Barnes, and Wolfgang Bangerth for all the stimulating discussions that took
place during the conception and implementation of the work at hand.
Furthermore, the authors gratefully acknowledge the support of Isuru
Fernando and Ray Donnelly during the implementation of the conda recipe.

\section*{Funding}

This work was supported by the German Research Foundation (DFG) within the
Heisenberg program (JI 115/4-2) and (in cooperation with the Technical
University of Munich) in the framework of the Open Access Publishing Program.

\bibliographystyle{osajnl}

\end{document}